\documentclass[lettersize,journal]{IEEEtran}
\usepackage{amsmath,amsfonts}
\usepackage{algorithm}
\usepackage{array}
\usepackage{textcomp}
\usepackage{stfloats}
\usepackage{url}
\usepackage[colorlinks=true, allcolors=blue]{hyperref}
\usepackage{verbatim}
\usepackage{graphicx}
\usepackage{cite} 
\usepackage{booktabs}
\usepackage{multicol}
\usepackage{graphicx}
\usepackage[export]{adjustbox}
\usepackage{array}
\usepackage{xcolor}
\usepackage{algorithm}
\usepackage{algpseudocode} 
\usepackage{amsmath}
\usepackage[caption=false,font=footnotesize]{subfig}

\newcolumntype{P}[1]{>{\centering\arraybackslash}p{#1}}

\begin{document}
\bstctlcite{IEEEexample:BSTcontrol}

\title{A Cylindrical Nanowire Array-Based Flexure-FET Receiver for Molecular Communication}

\author{Dilara~Aktas,~\IEEEmembership{Graduate Student Member,~IEEE,}
       Ozgur B.~Akan,~\IEEEmembership{Fellow,~IEEE}

\thanks{The authors are with the Center for neXt-generation Communications (CXC), Koç University, 34450 Istanbul, Turkey. O.B. Akan is also with the Internet of Everything (IoE) Group, Department of Engineering, University of Cambridge, CB3 0FA Cambridge, U.K. (e-mail: dilaraaktas20@ku.edu.tr, oba21@cam.ac.uk).}

\thanks{This work was supported in part by the AXA Research Fund (AXA Chair for Internet of Everything at Ko\c{c} University).}}
\markboth{}%
{}

\maketitle

\begin{abstract}

Molecular communication (MC) enables biocompatible and energy-efficient information transfer through chemical signaling, forming a foundational paradigm for emerging applications in the Internet of Nano Things (IoNT) and intrabody healthcare systems. The realization of this vision critically depends on developing advanced receiver architectures that merge nanoscale communication and networking techniques with bio-cyber interfaces, ensuring energy-efficient, reliable, and low-complexity modulation and detection while maintaining biocompatibility. To address these challenges, the Flexure-FET MC receiver was introduced as a mechanically transducing design capable of detecting both charged and neutral molecular species. In this study, we present a cylindrical nanowire array–based Flexure-FET MC receiver that enhances design versatility and scalability through distributed electromechanical coupling in a suspended-gate configuration. The proposed array architecture offers additional geometric degrees of freedom, including nanowire radius, length, spacing, and array size, providing a flexible framework that can be tailored to advanced MC scenarios. An analytical end-to-end model is developed to characterize the system’s electromechanical response, noise behavior, and information-theoretic performance, including signal-to-noise ratio (SNR) and channel capacity. The results reveal the strong interdependence between geometry, electromechanical dynamics, and molecular binding processes, enabling tunable control over sensitivity, noise characteristics, and communication capacity. The enhanced structural tunability and array configuration of the proposed design provide a flexible foundation for future mixture-based and spatially modulated MC systems, paving the way toward scalable and multifunctional receiver architectures within the IoNT framework.
\end{abstract}


\section{Introduction}

\IEEEPARstart{M}{olecular} communication (MC) is an emerging paradigm that conveys information between living entities and engineered systems through chemical signaling \cite{akyildiz2015internet,nakano2013molecular}. It governs information exchange among living organisms, from bacterial colonies to human cells, providing inherently biocompatible and energy-efficient signaling where conventional communication methods fail. Its inherent biocompatibility and energy efficiency make it a fundamental enabler of nanonetworks and the Internet of Everything (IoE) \cite{nakano2012molecular, akan2016fundamentals}. Ongoing progress in nanotechnology is enabling the conceptualization and early development of intrabody and body-area nanonetworks aimed at continuous and biocompatible physiological monitoring. These systems open new possibilities for next-generation medical applications, such as real-time health tracking, targeted drug delivery, disease diagnostics, and lab-on-chip devices, that can be developed based on information and communication technology (ICT) \cite{al2019internet, felicetti2016applications, kuscu2019transmitter}.

\begin{figure*}[!t]
\centering
\includegraphics[width=0.8\linewidth]{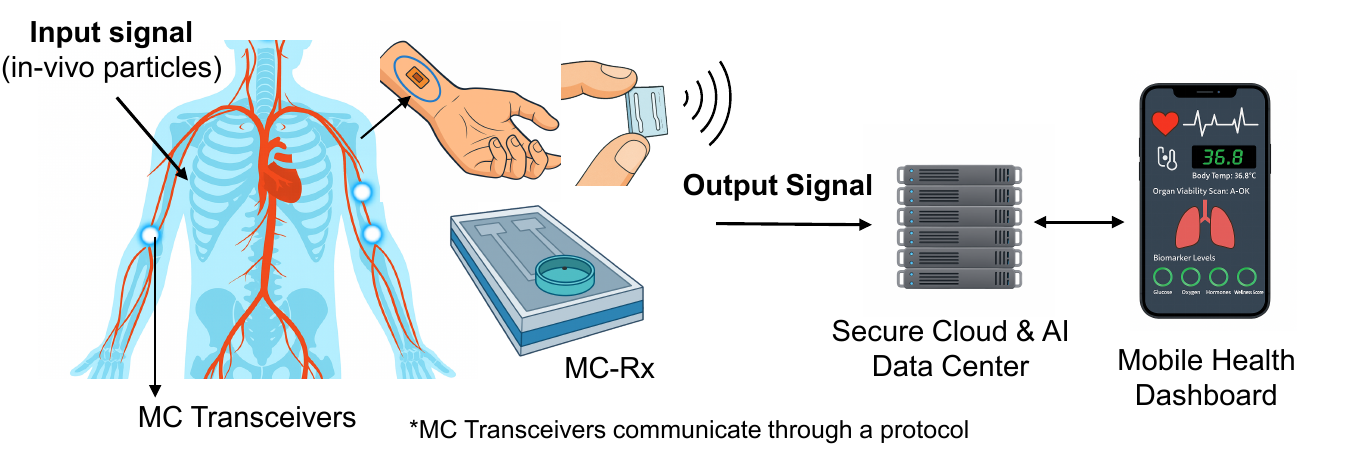}
\caption{Conceptual illustration of an IoNT-enabled healthcare system. In-vivo molecular signals are generated and transmitted by MC transceivers (MC-TxRxs), received by an MC-Rx device, and converted into electrical output signals \cite{koo2020deep}. These signals are processed through a secure cloud and AI data center, enabling real-time monitoring via a mobile health dashboard.}
\label{iont}
\end{figure*}

Within this landscape, intrabody healthcare applications naturally align with the Internet of Nano Things (IoNT) vision, which interconnects nanoscale devices and biological systems with larger cyber infrastructures to enable seamless bio-cyber communication and control \cite{akyildiz2015internet}. A schematic representation of such an IoNT-enabled healthcare network, where MC transceivers, i.e., MC-TxRxs, exchange biochemical signals that are converted into electrical outputs and processed through cloud-based platforms for real-time monitoring, is shown in Fig.~\ref{iont}. Achieving this vision requires efficient bio-cyber interfaces and biocompatible, energy-aware transceivers capable of linking biological environments with digital systems. Although substantial progress has been achieved in developing modulation, detection, and channel models, the engineering and physical realization of practical MC-TxRxs remain largely unexplored \cite{kuscu2016physical}. Bridging this gap requires practical transceiver architectures that embody the stochastic, nonlinear, and dynamic nature of molecular interactions, enabling the transition of MC  theory into functional IoNT systems \cite{kuscu2019transmitter}.
 
Recent research has therefore focused on the development of physical device models and bio-cyber interfaces that enable seamless integration between biological and artificial systems within the IoNT framework \cite{el2020mixing, chude2016biologically, akyildiz2019microbiome}. Early approaches predominantly relied on biologically derived nanomachines \cite{nakano2014molecular, unluturk2015genetically}, offering excellent biocompatibility but limited computational capacity and restricted applicability to \textit{in vivo} environments \cite{kuscu2016modeling}. To overcome these constraints, artificial structure-based MC receivers (MC-Rxs) have emerged as versatile alternatives, capable of continuous, label-free operation in both \textit{in vivo} and \textit{in vitro} conditions \cite{kuscu2016physical, kuscu2019transmitter}. Among various implementations, field-effect transistor (FET)–based biosensors, i.e., BioFETs, have attracted significant attention as MC-Rx architectures owing to their high sensitivity and seamless compatibility with nanoscale integration \cite{kuscu2016physical, kuscu2016modeling}. Nonetheless, their charge-based detection mechanism restricts the sensing of neutral molecules and degrades performance under high ionic concentrations, where Debye screening effects become dominant \cite{kuscu2016physical}.

\begin{figure}[!t]
\centering
\includegraphics[width=0.7\columnwidth]{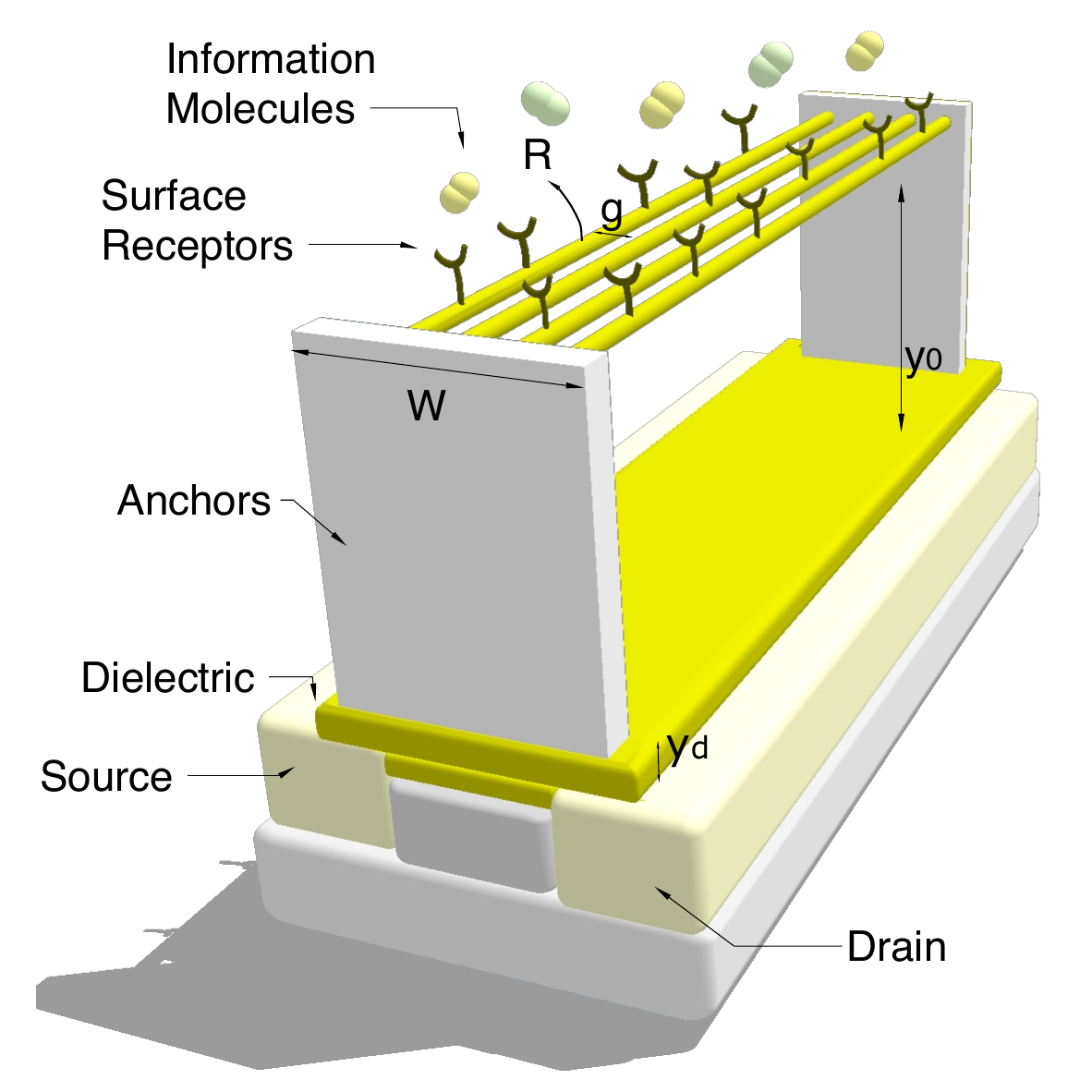}
\caption{Cylindrical nanowire array-based Flexure-FET MC-Rx, where $W$, $L$, and $R$ denote the beam width, length, and radius, respectively; $g$ is the inter-cylinder spacing, $y_0$ the initial air gap, and $y_d$ the dielectric thickness. Ligand binding on receptor-coated cylinders alters beam stiffness and deflection, modulating the output current.}
\label{fig2}
\end{figure}

To address these challenges, we developed the Flexure-FET MC-Rx \cite{aktas2021mechanical}, which employs mechanical transduction to detect both charged and neutral molecules with high sensitivity. Building on this foundation, we later introduced the first Weight Shift Keying (WSK)–based molecular communication (MC) system employing the Flexure-FET MC-Rx \cite{aktas2022weight} and analyzed its performance under biologically relevant interference conditions. In our most recent study \cite{aktas2025flexure}, the Flexure-FET MC-Rx framework was further advanced through the integration of a competitive binding model, enabling accurate characterization of multi-species interactions in realistic biological environments. By capturing receptor competition among coexisting molecular species, this extended framework improves receiver reliability and accuracy, enhancing the signal-to-noise ratio (SNR) and reducing the symbol error probability (SEP) in complex multitarget communication scenarios.

In this paper, we further advance the Flexure-FET–based MC framework by introducing a cylindrical nanowire array architecture, illustrated in Fig.~\ref{fig2}, that enhances sensitivity and scalability through a suspended-gate configuration. The proposed cylindrical nanowire array–based Flexure-FET MC-Rx, referred to as the cylindrical array–based Flexure-FET MC-Rx for short, introduces a distributed electromechanical sensing platform where multiple suspended cylindrical nanowires collectively function as a deformable gate electrode. This configuration captures the interplay among geometry, mechanical flexibility, and molecular binding dynamics, enabling a tunable electromechanical response and providing a physically grounded receiver design for MC systems.

We extend the previously established Flexure-FET formulation to analytically model the system’s electromechanical response, noise characteristics, and information-theoretic performance, including the SNR and channel capacity, constructing a comprehensive end-to-end framework. The results demonstrate how nanowire array geometry and receptor kinetics jointly determine molecular detection efficiency and communication reliability, laying the foundation for next-generation MC-Rx designs within the IoNT paradigm.

Building upon the groundwork established in \cite{aktas2025flexure}, which advanced the Flexure-FET MC-Rx concept toward the development of future odor-based molecular communication (OMC) receivers through chemical interaction and interference modeling, this study furthers that vision from a geometrical and physical design perspective. The proposed cylindrical array–based Flexure-FET MC-Rx introduces distributed electromechanical coupling and geometry-driven tunability, forming the structural foundation for next-generation OMC receivers (OMC-Rxs) \cite{aktas2024odor}. Each nanowire within the array can be independently functionalized with distinct receptor types, enabling spatially resolved and receptor-selective molecular recognition. This configuration allows simultaneous transduction of multiple binding events, supporting mixture decomposition and multi-ligand detection in complex molecular environments. Moreover, the spatial organization of the nanowires can enable spatially modulated MC schemes, where information is encoded not only in ligand type or concentration but also through the selective activation of specific nanowires. Collectively, these advances bridge the gap between theoretical OMC frameworks and physically realizable receiver architectures, paving the way for the practical deployment of OMC systems within the IoE paradigm.

The rest of the paper is organized as follows: Section II introduces the Cylindrical Nanowire Array-Based Flexure-FET Molecular Communication Receiver, describing its operating principles and distributed electromechanical model. Section III presents the MC System Model, including signal flow, biorecognition, transduction, and noise characterization. Section IV provides the Numerical Evaluation, covering sensitivity, receiver noise, SNR, and capacity analysis. Finally, Section V concludes the paper.

\begin{figure}[!t]
\centering
\includegraphics[width=0.8\linewidth]{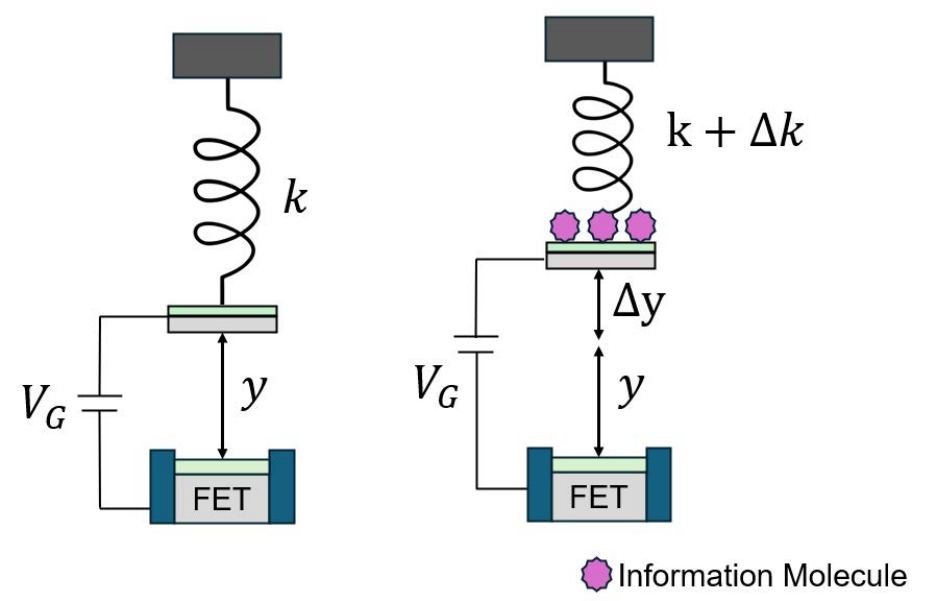}
 \caption{Spring–mass model representation of the Flexure-FET. The suspended gate is modeled as a spring with stiffness $k$, which increases to $k + \Delta k$ upon ligand binding. The resulting deflection change $\Delta y$ under a constant gate bias $V_G$ alters the air gap, thereby modulating the channel current through mechanical-to-electrical transduction.}
    \label{fig:flexurefet_spring_model}
\end{figure}

\section{Cylindrical Nanowire Array-Based Flexure-FET Molecular Communication Receiver}

In this section, the operating principles of the cylindrical nanowire array-based Flexure-FET MC-Rx are described, along with the main factors influencing its sensitivity.  
In particular, the effects of array geometry, inter-wire spacing, and material properties on the device behavior are analyzed in detail through the electrostatic and electromechanical formulations presented in the subsequent subsections.  
These analyses establish the pre-capture (i.e., before the binding of target molecules) electromechanical state of the receiver, which serves as the foundation for modeling post-capture variations and current modulation, as will be detailed in Section III-B.

\subsection{Operating Principles}

The structural and functional configuration of the receiver is illustrated in Fig.~\ref{fig2}.  
Ligands bind to receptors functionalized on the outer surfaces of cylindrical nanowires suspended above the transistor channel.  
Each nanowire is fixed at both ends, forming a fixed–fixed beam configuration that is periodically repeated to create an aligned array.  
Ligand adsorption generates surface stress, which modifies the effective stiffness of the nanowires and shifts their equilibrium position. 
This change in mechanical state alters the air gap between the nanowires and the gate dielectric, thereby modulating the effective capacitance of the system, as shown in Fig.~\ref{fig:flexurefet_spring_model}.  
These electromechanical variations are transduced into changes in surface potential and drain current, forming the basis of the Flexure-FET’s sensing mechanism \cite{jain2012flexure}.

Flexure-FET operation is governed by two coupled nonlinear mechanisms: \textit{spring softening} and \textit{subthreshold electrical conduction} \cite{jain2012flexure}.  
In the \textit{spring-softening} regime, the effective stiffness decreases nonlinearly with increasing gate bias \(V_G\) until the onset of pull-in instability, where the structure becomes mechanically unstable.  
The device is biased close to this critical point to achieve maximum sensitivity.  
In the \textit{subthreshold} region, the exponential dependence of drain current on surface potential further amplifies the device response to small perturbations.  
The strong coupling between the mechanical and electrical responses allows the Flexure-FET to achieve high transduction gain with minimal power consumption.  
Its operation is inherently reference-free, simplifying on-chip integration.  
Since the transduction mechanism is based on stiffness modulation rather than charge transfer, the device is capable of detecting both charged and neutral biomolecules, enabling reliable and selective sensing for MC applications.

\subsection{Electromechanical Model}

In \cite{aktas2021mechanical, aktas2022weight}, the Flexure-FET was implemented as a mechanical MC-Rx using a single-beam configuration, adapted from the biosensor model originally introduced in \cite{jain2012flexure}.  
This implementation demonstrated enhanced sensitivity by operating near the pull-in instability and within the subthreshold conduction regime.  
In the present study, the concept is extended to an array-based cylindrical electrode geometry to further improve sensitivity and stability through distributed electromechanical coupling.

\subsubsection*{Mechanical Behavior of the Nanowire Array}

The array-based configuration introduces distributed mechanical coupling among multiple suspended elements, requiring an accurate representation of each beam’s elastic behavior. Each cylindrical microbeam is modeled as a fixed–fixed spring element with stiffness,
\begin{equation}\label{eq:1}
k_{\text{single}} = \frac{\alpha\,E\,\pi R^4}{4L^3},
\end{equation}
where \(E\) is the Young’s modulus, \(R\) is the cylinder radius, \(L\) is the beam length, and \(\alpha\) is a geometric factor.
For an array of \(N_{\text{array}}\) identical linear springs operating in parallel, the effective stiffness is expressed as,
\begin{equation}\label{eq:2}
k_{\text{array}} = N_{\text{array}}\,k_{\text{single}}.
\end{equation}

The equilibrium position of the gate is governed by the balance between the mechanical restoring force and the electrostatic attraction, \(F_s = F_{\text{elec}}\).  
The electrostatic force depends on the array electrode geometry and the inter-cylinder spacing \(g\), which determines the degree of electrostatic coupling among adjacent cylinders.  
For small spacing, the array behaves as a planar electrode, while for large spacing, it approaches the single-cylinder limit.  
These geometric parameters strongly affect device stability and pull-in characteristics \cite{thjain2014fundamental}. At the onset of pull-in instability, the balance between electrostatic and elastic forces defines the critical gap ratio of the structure.  
Biasing the device close to this point enhances mechanical sensitivity while maintaining stable electromechanical operation.

\subsubsection*{Electrostatic Characteristics of the Array Electrode}

The electrostatic behavior of the cylindrical array-based Flexure-FET differs from planar and single-cylinder electrodes because the electric field distribution is influenced by the spacing between adjacent nanowires, which modifies the overall capacitance and electrostatic force.
For the array geometry, the effective capacitance is expressed as \cite{thjain2014fundamental},
\begin{equation}\label{eq:carr}
C_{\text{array}} = 
\frac{2\pi\varepsilon_0 L}
{\ln\!\left(\dfrac{\sinh\!\left(\tfrac{2\pi(y+R)}{g}\right)}{\tfrac{\pi R}{g}}\right)},
\end{equation}
where \(L\) is the nanowire length, \(R\) is the nanowire radius, and \(g\) denotes the center-to-center spacing between adjacent nanowires in the array.  
These geometric parameters jointly define the array coupling strength and govern the transition between the planar and cylindrical electrostatic regimes.  
Specifically, the expression in (\ref{eq:carr}) reduces to the cylindrical form \(C_{\text{cyl}} = 2\pi\varepsilon_0 L / \ln(2(1+y/R))\) for large inter-wire spacing (\(g \to \infty\)), and to the planar capacitance \(C_{\text{pl}} = \varepsilon_0A_e/y\) when (\(g \to 0\)).  Accordingly, the electrostatic force for voltage actuation in the array configuration is given by \cite{thjain2014fundamental},
\begin{equation}\label{eq:4}
F_{\text{elec,array}} =
\frac{2\pi^2 \varepsilon_0 L \coth\!\left(\tfrac{2\pi(y+R)}{g}\right)}
{g\!\left[\ln\!\left(\dfrac{\sinh\!\left(\tfrac{2\pi(y+R)}{g}\right)}{\tfrac{\pi R}{g}}\right)\right]^2}
V_G^2.
\end{equation}
  
These expressions highlight the strong dependence of capacitance and electrostatic coupling on the array spacing, enabling tunable sensitivity and reduced pull-in voltage through geometric optimization. Unlike the single-beam or planar Flexure-FET designs used in earlier M architectures, the proposed array configuration introduces additional design flexibility by allowing simultaneous tuning of multiple structural parameters, including the nanowire radius \(R\), length \(L\), inter-wire spacing \(g\), and number of nanowires \(N_{\text{array}}\).  
This multidimensional control enables a broader optimization space for achieving the desired electromechanical response by balancing sensitivity, stability, and operating voltage.

Moreover, the array geometry inherently supports greater mechanical compliance and flexible surface functionalization, where each nanowire can be selectively coated with receptor or sensing layers for enhanced selectivity or multi-analyte detection.  
The cylindrical array design is also compatible with standard MEMS and nanowire fabrication processes such as SU-8-based lithography and template-assisted growth, enabling scalable and reproducible device realization \cite{fruncillo2021lithographic}. In addition, SU-8 serves as an excellent structural material due to its low Young’s modulus, which reduces the overall stiffness of the beams and thereby enhances the electromechanical sensitivity of the receiver.  

\subsubsection*{Self-Consistent Solution for System State}

The coupled electromechanical behavior of the cylindrical array-based Flexure-FET is determined by the balance between mechanical restoring and electrostatic forces, which are solved self-consistently:
\begin{align}
F_s &= k_{\text{array}}(y_0 - y) = F_{\text{elec,array}}(y, V_G), \label{eq:eq1}\\[4pt]
V_G &= (y + y_d/\varepsilon_d)\,\varepsilon_s E_s(\psi_s) + \psi_s, \label{eq:eq2}
\end{align}
where  \(y_0\) is the initial air gap, and \(\psi_s\) is the surface potential.  
The first equation expresses mechanical equilibrium, while the second relates \(V_G\) to the electric field at the semiconductor interface and the potential drops across the dielectric and air gaps. The electric field below the membrane, \(E_{\text{air}}\), is equal to \(\varepsilon_s E_s(\psi_s)\), where \(\varepsilon_s\) is the dielectric constant of the substrate, and \cite{jain2012flexure},
\begin{equation}
\begin{aligned}
E_s(\psi_s)= &
\sqrt{\frac{2 q N_A}{\varepsilon_0 \varepsilon_s}}
\left[
\psi_s+\left(e^{\frac{-q \psi_s}{k_B T}}-1\right)\frac{k_B T}{q}\right. \\
& \left.
-\left(\frac{n_i}{N_A}\right)^2
\left(
\psi_s-\left(e^{\frac{q \psi_s}{k_B T}}-1\right)\frac{k_B T}{q}
\right)
\right]^{\frac{1}{2}},
\end{aligned}
\end{equation}
where \(E_s(\psi_s)\) is the electric field at the substrate–dielectric interface, \(q\) is the electron charge, \(N_A\) is the substrate doping concentration, \(k_B\) is the Boltzmann constant, \(T\) is the absolute temperature, and \(n_i\) is the intrinsic carrier concentration in the substrate.  
The voltage drop in the air gap \((y\,\varepsilon_s E_s(\psi_s))\), dielectric $\left(\frac{y_d}{\epsilon_d} \epsilon_s E_s\left(\psi_s\right)\right),$
 and substrate \((\psi_s)\) is expressed in terms of the applied gate voltage as given in (\ref{eq:eq2}).  
Accordingly, \(y\) (the position
of the gate electrode) and \(\psi_s\) are obtained by solving the coupled relations self-consistently.  
This analysis yields the steady-state deflection \(y\), potential \(\psi_s\), and capacitance \(C_{\text{array}}\) as functions of \(V_G\), defining the pre-capture state from which post-capture variations \((\Delta k,\, \Delta \psi_s,\, \Delta y)\) and current modulation are derived in subsequent sections.

\begin{figure}[!t]
\centering
\includegraphics[width=\linewidth]{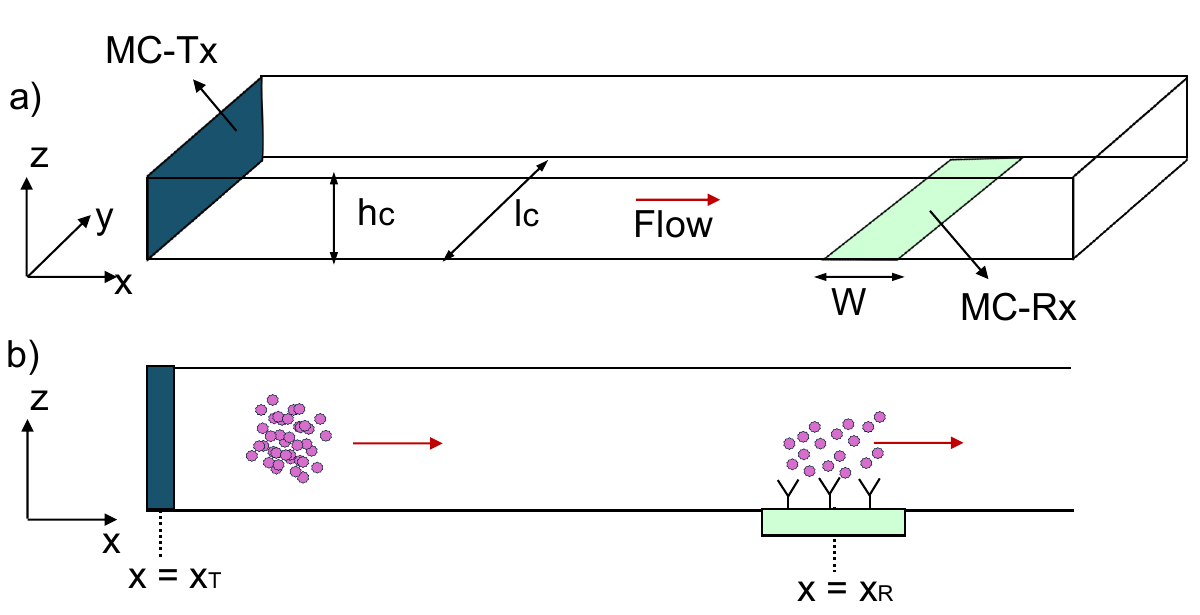}
\caption{(a) 3D and (b) 2D views of a microfluidic channel \cite{kuscu2016modeling}.}
\label{fig4}
\end{figure}

\begin{figure*}[!t]
\centering
\includegraphics[width=0.9\linewidth]{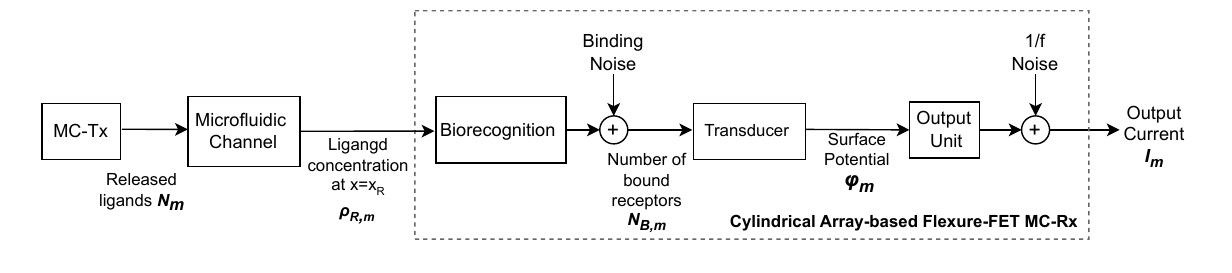}
\caption{The block diagram of the microfluidic MC system \cite{kuscu2016modeling}.}
\label{fig5}
\end{figure*}

\section{MC System Model}

We consider a diffusion–advection-based MC system comprising a single transmitter–receiver pair within a confined microfluidic environment.  
The channel is modeled as a rectangular channel that represents physiological pathways relevant to intrabody or body-area nanonetworks.  
The planar transmitter, positioned at \(x = x_T\), functions as a controlled molecular source that releases ligand molecules into the flowing medium.  
Downstream, the cylindrical array-based Flexure-FET MC-Rx is integrated at the channel substrate at \(x = x_R\), positioned to minimize hydrodynamic disturbances while maintaining efficient molecular capture. A schematic of the system geometry and flow topology is shown in Fig.~\ref{fig4}.

At the beginning of each signaling interval \(kT_s\), a fixed number of ligands \(N_m\) are released into the microfluidic channel \cite{kuscu2016modeling}.  
These ligands propagate through diffusion and advection toward the receiver, where they interact with surface-bound receptors on the MC-Rx.  
The stochastic binding–unbinding events that occur at the receptor interface can be statistically described by the mean and variance of the number of bound receptors, capturing the instantaneous molecular reception behavior.  
These statistical quantities form the foundation for analyzing the receiver response, SNR, and information transfer characteristics.  
The functional blocks of the overall microfluidic MC system, including molecular propagation, biorecognition, and electromechanical transduction, are illustrated in Fig.~\ref{fig5}.

\subsection{Signal Flow \& Biorecognition}

The propagation of signaling molecules through the microfluidic channel and their subsequent interaction with the receiver surface can be modeled as a coupled advection–diffusion and mass-transfer process. Under steady laminar flow, molecular dispersion follows the Taylor–Aris mechanism \cite{dutta2006effect}, where the effective diffusion coefficient for a rectangular microchannel is expressed as \cite{bicen2013system},
\begin{equation}\label{eq:2}
D = \Biggl(1 + \frac{8.5u^2h_c^2l_c^2}{210D_0^2(h_c^2 + 2.4h_cl_c + l_c^2)}\Biggr) D_0,
\end{equation}
with $u$ denoting the mean flow velocity, and $h_c$ and $l_c$ the height and width of the channel cross-section, respectively.  

The spatial ligand concentration $\rho_m(x,t)$ follows a Gaussian distribution for $t>0$, which is the solution of the one-dimensional advection–diffusion equation along the flow direction $x$, assuming a single signaling interval and neglecting intersymbol interference (ISI):
\begin{equation}\label{eq:3}
\rho_m(x,t) = \frac{N_m/A_c}{\sqrt{4\pi D t}}
\exp\!\Biggl(-\frac{(x - ut)^2}{4Dt}\Biggr),
\end{equation}
where $A_c = h_c \times l_c$ is the cross-sectional area of the channel. Ligands released at $t=0$ are assumed to be uniformly distributed across $A_c$, i.e., $\rho_m(x,0) = (N_m/A_c)\delta(x)$ \cite{kuscu2016modeling}.

At the receiver located at $x_R$, the peak of the molecular pulse reaches the surface after a propagation delay given by,
\begin{equation}\label{eq:4}
t_D = \frac{x_R}{u}.
\end{equation}

The instantaneous ligand concentration sampled by the receiver at the sampling time $t_D$ can be approximated by the peak value of the molecular pulse, i.e., $\rho_m(x_R,t_D) = \rho_{R,m}$, 
\begin{equation}\label{eq:5}
    \rho_m(x_R,t) \approx \rho_m(x_R,t_D) = \frac{N_m}{A_{c}\sqrt{4\pi Dt_D}}, 
\end{equation}
for $t \in [t_D - \tau_p/2,\, t_D + \tau_p/2]$, where $x_R$ denotes the receiver center position and $\tau_p$ represents the approximate duration of the ligand passage across the receiver surface \cite{kuscu2016modeling}.  

The proportional relationship between the transmitted and received ligand concentrations can be expressed through the channel constant,
\begin{equation}\label{eq:cons}
\beta_{\text{ch}} = \frac{1}{A_c\sqrt{4\pi D t_D}},
\end{equation}
which quantifies the molecular transport efficiency in the advection–diffusion channel as a function of flow velocity, diffusion coefficient, and channel geometry.

At the receiver interface, ligand transfer from the bulk flow to the active surface of the cylindrical nanowire array is quantified by the dimensionless parameter,

\begin{equation}\label{eq:7}
P_s = \frac{6Q(w_R^{\mathrm{eff}})^2}{D\,l_c\,h_c^2},
\end{equation}
where $Q = u \times A_c$ is the volumetric flow rate and $w_R^{\mathrm{eff}}$ denotes the effective width of the cylindrical nanowire array.

For a rectangular cross-section microfluidic channel, the corresponding surface mass-transfer coefficient \(k_T\), which characterizes the transport rate (i.e., flux) toward the receiver surface, is given by \cite{zhang1996mass}:

\begin{equation}\label{eq:8}
{k_T} =  Dl_c \times 
\begin{cases}
(0.8075P_s^{1/3} + 0.7058P_s^{-1/6} 0.1984P_s^{-1/3}), \\ 
{\text{ if}}\ {P_s > 1} \\ 
{\frac{2\pi}{4.885 - \ln{(P_s)}}(1-\frac{0.09266P_s}{4.885 - \ln{(P_s)}})},\\ 
{\text{ if}}\ {P_s < 1} \\ 
\end{cases}
\end{equation}
For the biorecognition modeling, the interaction between transported ligands and the receptor-coated receiver surface is described using affinity-based recognition as the fundamental sensing principle~\cite{rogers2000principles}. In this mechanism, ligands transiently attach to receptor sites and subsequently dissociate, enabling a reversible and dynamic detection process~\cite{kuscu2016modeling}. Under steady-state and equilibrium conditions, the mean number of bound receptors on the receiver surface is expressed as follows \cite{kuscu2016modeling},

\begin{equation}\label{eq:6}
    \mu_{N_{B,m}} = P_{\text{on}|m}N_R = \frac{\rho_{R,m}}{\rho_{R,m}+ K_D}N_R,
\end{equation}
where $P_{\text{on}|m}$ represents the steady-state probability that a receptor is occupied by a ligand (ON state), 
$K_D = k_{-1}/k_1$ denotes the dissociation constant, and $k_1$ and $k_{-1}$ correspond to the effective association and dissociation rate constants, respectively~\cite{rogers2000principles,berezhkovskii2013effect}. 
The total number of surface receptors is given by 
$N_R = \rho_{SR} A_{\text{eff}}$, where $\rho_{SR}$ is the receptor surface density and $A_{\text{eff}}$ is the effective surface area of the array ($A_{\text{eff}} = A$). 
The number of bound receptors statistically follows a binomial distribution, whose variance can be expressed as~\cite{kuscu2016modeling},

\begin{equation}\label{eq:7}
    \sigma^2_{N_{B,m}} = P_{\text{on}|m}(1-P_{\text{on}|m})N_R.
\end{equation}

These analytical expressions for the mean and variance of bound receptors remain valid for the cylindrical array-based Flexure-FET MC receiver, as the ligand concentration is assumed to be uniformly distributed over the active surface and each receptor acts independently, similar to the planar and single-nanowire cases.

\subsection{Transducer}

Flexure-FET transduction relies on three sequential processes: the modification of mechanical stiffness \(\Delta k\) induced by ligand binding, the gate displacement \(\Delta y\) under near pull-in bias conditions, and the exponential response of drain current in the subthreshold regime \cite{jain2012flexure}.

The mean density of bound molecules can be defined using (\ref{eq:6}) as $\mu_{N_s} = \mu_{N_{B,m}}A^{-1}$ where $N_s$ is the bound ligand density at the receiver surface. The corresponding mean variations in stiffness \(\mu_{\Delta k_m}\), deflection \(\mu_{\Delta y_m}\), surface potential \(\mu_{\Delta \psi_m}\), and the mean current ratio (sensitivity) \(\mu_{S_m}\) for symbol \(m\) can then be determined.

For the cylindrical nanowire configuration, the stiffness of a single beam is given by (\ref{eq:1}) as \(k_{\text{single}} \propto R^4\).  
A small radial increase due to ligand binding, \(R \rightarrow R + \Delta R\), modifies the stiffness according to,  
\begin{equation}
\frac{\Delta k_{\text{single}}}{k_{\text{single}}} 
\approx 4\,\frac{\Delta R}{R},
\end{equation}
where the coefficient arises from the quartic dependence of stiffness on the nanowire radius.  
Assuming volume conservation, the effective radial change can be expressed as,  
\begin{equation}
\Delta R = \mu_{N_s} A_t H_t,
\end{equation}
where \(A_t = \pi R_t^2\) is the molecular cross-sectional area and \(R_t\) is the radius of the target molecule.  
Substituting into the previous expression gives,  
\begin{equation}\label{eq:delta_k_single}
\frac{\mu_{\Delta k_{m,\text{single}}}}{k_{\text{single}}} 
\approx 4\,\mu_{N_s} A_t \frac{H_t}{R}.
\end{equation}
Finally, for an array of \(N_{\text{array}}\) nanowires operating in parallel, the mean effective stiffness change is obtained as,  
\begin{equation}\label{eq:delta_k_eff}
\mu_{\Delta k_{m,\text{eff}}} = N_{\text{array}}\,\mu_{\Delta k_{m,\text{single}}}.
\end{equation}
For bias close to pull-in, the mean change in gate position $\mu_{\Delta y_m}$ can be given as,

\begin{equation}
  \mu_{\Delta y_m}  \approx \sqrt{\frac{\epsilon_0A(V_G-\psi_s)^2}{2(3y-y_0)}\frac{\mu_{\Delta k_{m,\text{eff}}}}{k_{eff}^2}},
\end{equation}
where $\epsilon_0$ is the permittivity of free space, $\psi_s$ is the surface potential. The corresponding mean change in surface potential $\mu_{\Delta \psi_{s,m}} =\mu_{\Delta \psi_m}$ can be described as,

\begin{equation}\label{eq:14}
  \mu_{\Delta \psi_m} \approx \frac{-k \mu_{\Delta y_m} + \mu_{\mu_{\Delta k_{m,\text{eff}}}}(y_0 - y)}{q \epsilon_sN_AA},
\end{equation}
where $q$ is the charge of an electron, $\epsilon_s$ is the dielectric constant of the substrate, and $N_A$ is the substrate doping. In terms of the change in surface potential, the ratio of the drain current before $(I_{DS1})$, which is derived in \cite{jain2012flexure}, and after $(I_{DS2})$ capture refers to the mean sensitivity of the device $ S = \mu_{S_m}$ and can be expressed as,

\begin{equation}\label{eq:15}
    S = \frac{I_{DS1}}{I_{DS2}}  \approx \exp\biggl(\frac{k {\Delta y_m} - \mu_{\Delta k_{m,\text{eff}}}(y_0 - y)}{k_BT\epsilon_sN_AA}\biggl),
\end{equation}
where $I_{DS2} = \mu_{I_m} $ is the mean output current, $k_B$ is the Boltzmann constant, and $T$ is the absolute temperature.

\subsection{Noise}

The total noise affecting the Flexure-FET MC-Rx arises from multiple stochastic sources inherent to molecular binding and device operation.  
Among these, binding fluctuations and flicker noise dominate the output variance, while thermomechanical and temperature-induced effects are comparatively minor and thus neglected in the analysis. At steady state, binding noise arises from stochastic fluctuations in receptor occupancy, governed by reaction kinetics and modulated by the ligand transport rate \(k_T\).  
In the high \(k_T\) (reaction-limited) regime, the power spectral density (PSD) of the binding noise is expressed as~\cite{kuscu2016modeling},

\begin{equation}
S_{N_{B,m}}(f) = \sigma_{N_{B,m}}^2 \frac{2\tau_{B,m}}{1 + (2\pi f\tau_{B,m})^2},
\end{equation}
where $\tau_{B,m}$ is the relaxation time required for the receptor population to reach equilibrium, given by $\tau_{B,m} = 1 / (k_1 \rho_{R,m} + k_{-1})$.  

The flicker ($1/f$) noise current-referred power spectral density (PSD) is given by~\cite{rajan2010temperature},
\begin{equation}\label{eq:16}
S_{I_m^{F}}(f) = 
\frac{\lambda k_B T q^2 N_{ot} g_{\text{FET}}^2}
{w_R l_R C_{ox}^2 |f|}
\left[1 + \alpha_s \mu_p C_{ox} (V_G - |V_{TH}|)\right]^2,
\end{equation}
where $\lambda$ is the tunneling distance, $N_{ot}$ the oxide trap density,  
$g_{\text{FET}} = \partial I_{DS}/\partial V_G$ the transconductance,  
$\alpha_s$ the Coulomb scattering coefficient, $\mu_p$ the carrier mobility,  
$C_{ox}$ the oxide capacitance, and $V_{TH}$ the threshold voltage. The total current noise is modeled as the superposition of independent Gaussian processes representing binding and flicker noise~\cite{hooge1969amplitude}:
\begin{equation}
S_{I_m}(f) = S_{I_m^{B}}(f) + S_{I_m^{F}}(f),
\end{equation}
where  $S_{I_{B,m}}(f) =  S_{N_{B,m}}(f) \: \times \: \Delta \psi^2  \: \times \: g_{FET}^2 $, $\Delta \psi$ refers to the change in surface potential for a single bound ligand.  
The overall current variance is obtained as \cite{kuscu2016modeling},
\begin{equation}\label{eq:var}
\sigma_{I_m}^2 = \int_{-\infty}^{\infty} S_{I_m}(f)\, df.
\end{equation}

\section{Numerical Evaluation}

In this section, we present the numerical results obtained based on the analytical model introduced in the previous section to evaluate the performance of the cylindrical array-based  Flexure-FET MC-Rx. The analyses include the receiver sensitivity, individual noise components, signal-to-noise ratio (SNR), and channel capacity under various system configurations. Default parameter values for the simulations are provided in Table~\ref{tab:params}, selected in accordance with representative values commonly adopted in MC literature \cite{kuscu2016physical}.

\begin{figure}[!t]
  \centering
  \includegraphics[width=0.9\linewidth]{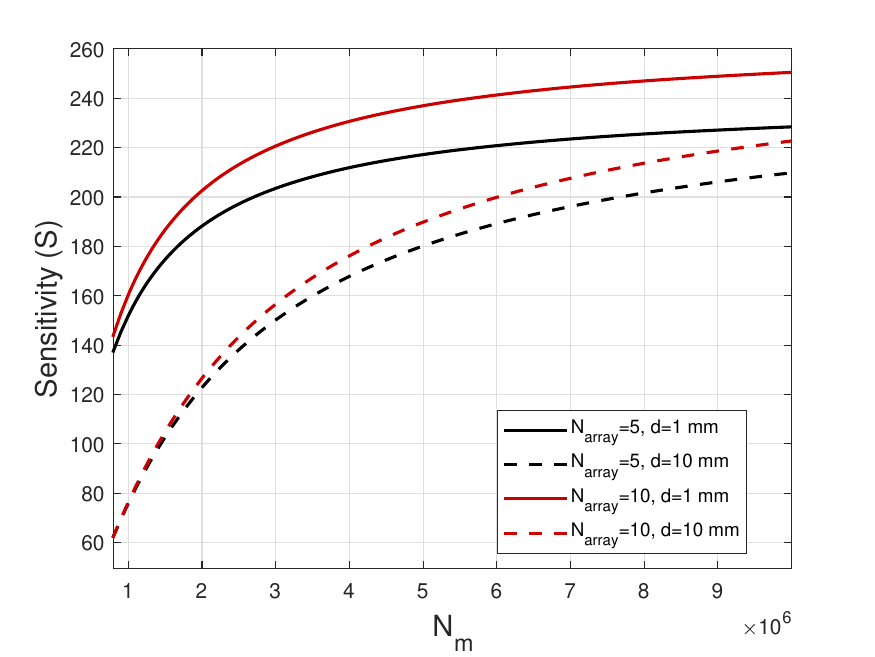}
\caption{Sensitivity $S$ of the cylindrical array-based Flexure-FET MC-Rx versus number of transmitted ligands $N_m$ for different transmitter--receiver distances ($d = 1$ and $10~\text{mm}$). Black and red curves represent $N_{\text{array}} = 5$ and $10$, respectively; solid and dashed lines correspond to $d = 1$ and $10~\text{mm}$.}

    \label{fig:sensitivity_array}
\end{figure}

\begin{table}
  \caption{Simulation Parameters}
  \label{tab:params}
  \begin{tabular}{ll}
    \toprule
    \midrule
    
Microfluidic channel length $(l_{c})$          & $4   \:  \mu m$\\
Microfluidic channel height $(h_{c})$          & $3   \:  \mu m$\\
Transmitter receiver distance $(d)$          & $10    \:  mm$\\
Number of transmitted ligands  $(N_m)$          & $1 \times10^{9} $\\
Average flow velocity  $(u)$          & $10   \:  \mu m/s$\\
Concentration of surface receptors  $(\rho_{SR})$ & $5 \times 10^{18}\:m^{-2} $\\
Binding rate  $(k_1)$  & $3 \times 10^{-16} \:  m^{3}/s $\\
Unbinding rate  $(k_{-1})$  & $20\:  s^{-1}  $\\
Young modulus of beam  $(E)$  & $4    \:  GPa$\\
Nanowire length $(L)$ & $4~\mu\text{m}$ \\
    Nanowire radius $(R)$ & $25~\text{nm}$ \\
Beam thickness $(H)$          & $260    \:  nm $\\
Air gap $(y_0)$          & $100    \:  nm $\\
Substrate doping $(N_A)$          & $10^{16}    \:  cm^{-3} $\\
Dielectric thickness $(y_d)$          & $5   \:  nm $\\
Oxide trap density  $(N_{ot})$  & $2.3 \times 10^{24} \:  eV^{-1}cm^{-3} $\\
  \bottomrule
\end{tabular}
\end{table}
\subsection{Sensitivity}

The expected sensitivity of the cylindrical array-based Flexure-FET MC-Rx is analyzed as a function of the number of transmitted ligands $N_m$, the transmitter–receiver distance $d$, and the number of nanowires in the receiver array $N_{\mathrm{array}}$. 

As shown in Fig.~\ref{fig:sensitivity_array}, the sensitivity, defined by the mean output current (i.e., $S = \mu_{S_m}$), increases with the number of ligands $N_m$ due to the enhanced molecular binding activity at the receiver surface. However, increasing the transmitter–receiver distance $d$ leads to higher molecular attenuation and consequently lower sensitivity because of the reduced ligand concentration at the receiver. Furthermore, a larger number of nanowires in the receiver array ($N_{\mathrm{array}}$) improves the effective surface area and total deflection, which enhances the overall transduction efficiency. 

\begin{figure}[!t]
  \centering
  \includegraphics[width=0.9\linewidth]{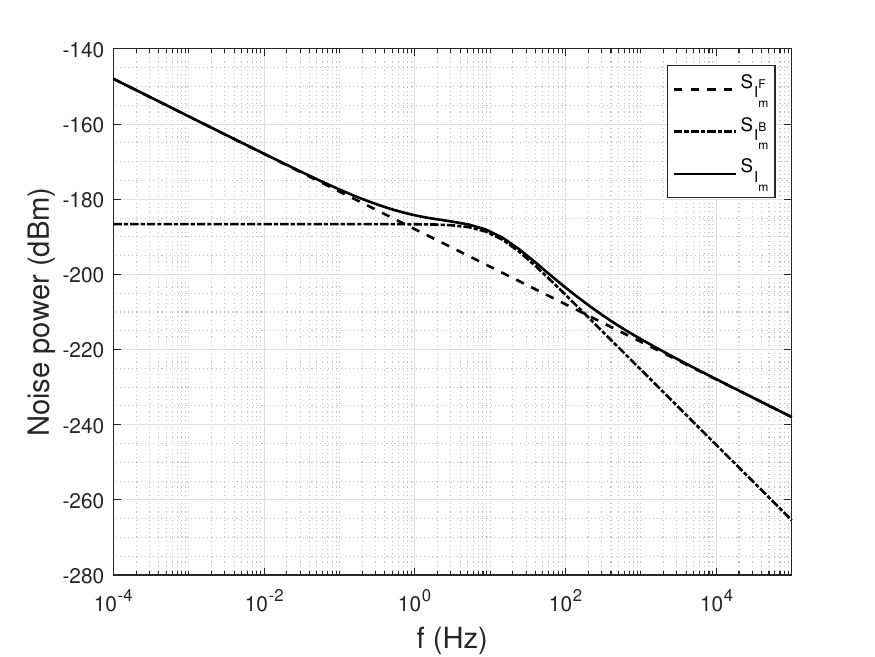}
\caption{Individual contributions of the noise sources in the cylindrical array-based Flexure-FET MC-Rx. 
    Here, $S_{I_{B,m}}$ denotes the binding noise, $S_{I_{F,m}}$ represents the flicker noise, 
    and $S_{I_m}$ corresponds to the power spectral density (PSD) of the total output current noise.}
  \label{noise}
\end{figure}
\begin{figure}[!t]
  \centering
  \includegraphics[width=0.9\linewidth]{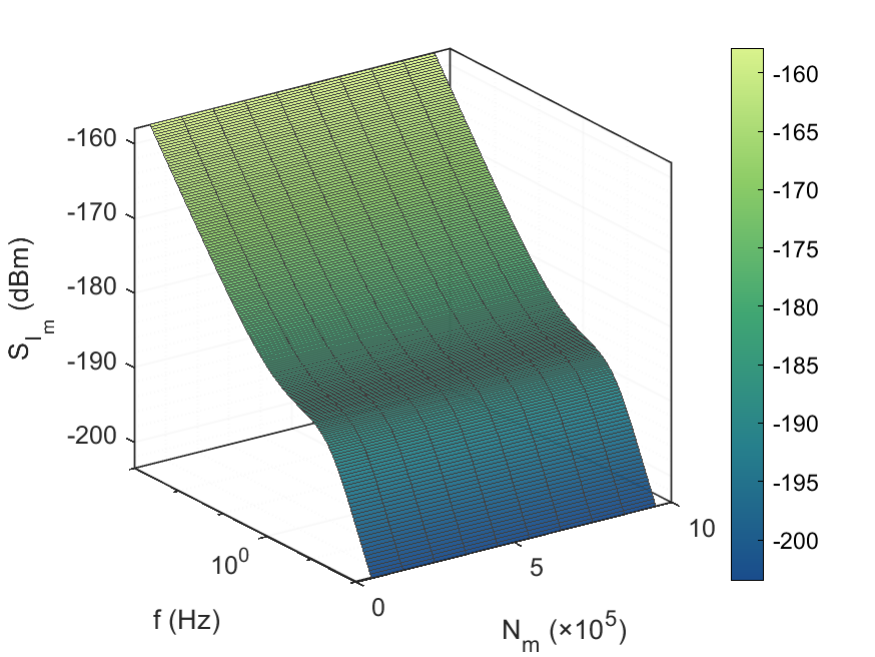}
\caption{The overall noise PSD of the array-based cylindrical Flexure-FET MC-Rx as a function of frequency $f$ and number of transmitted ligands $N_m$.}
  \label{noise3d}
\end{figure}

\begin{figure*}[!t]
\centering
\subfloat[]{\includegraphics[width=0.24\textwidth]{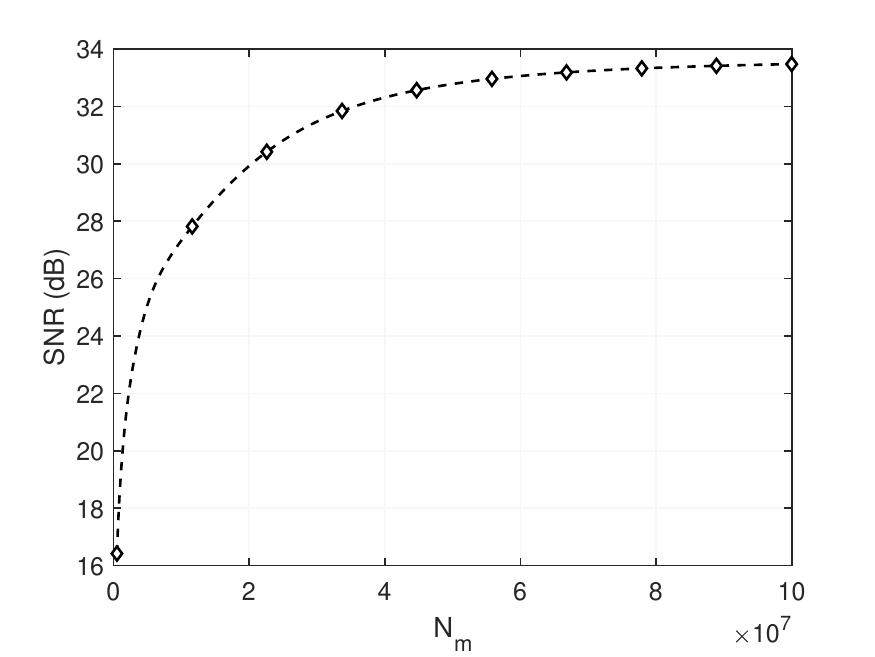}}
\hfil
\subfloat[]{\includegraphics[width=0.24\textwidth]{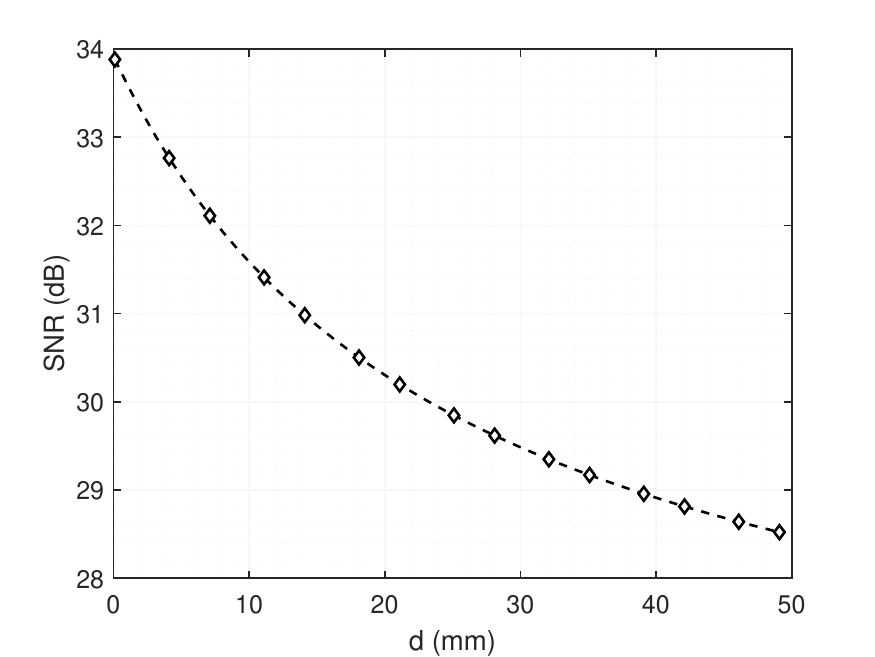}}
\hfil
\subfloat[]{\includegraphics[width=0.24\textwidth]{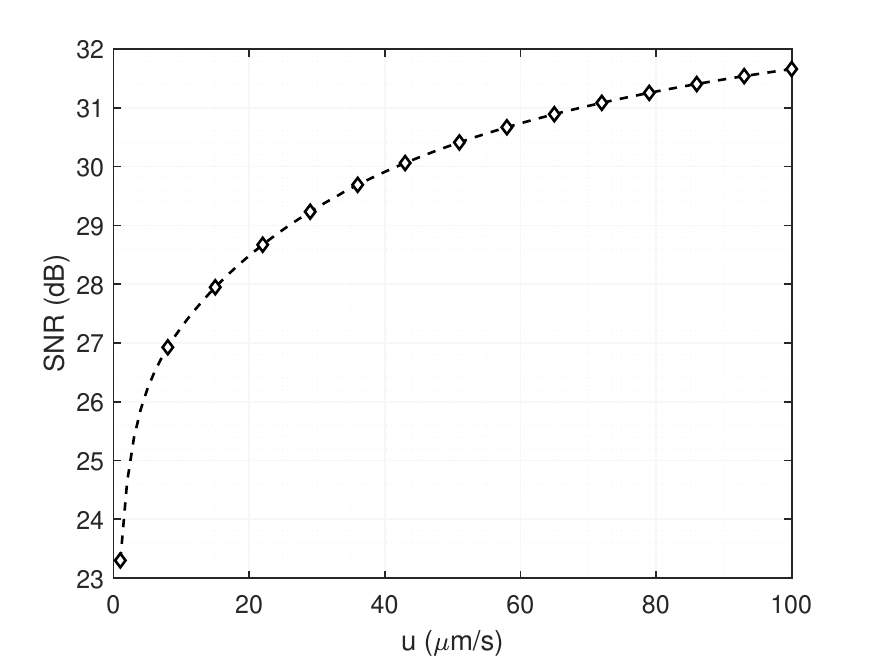}}
\hfil
\subfloat[]{\includegraphics[width=0.24\textwidth]{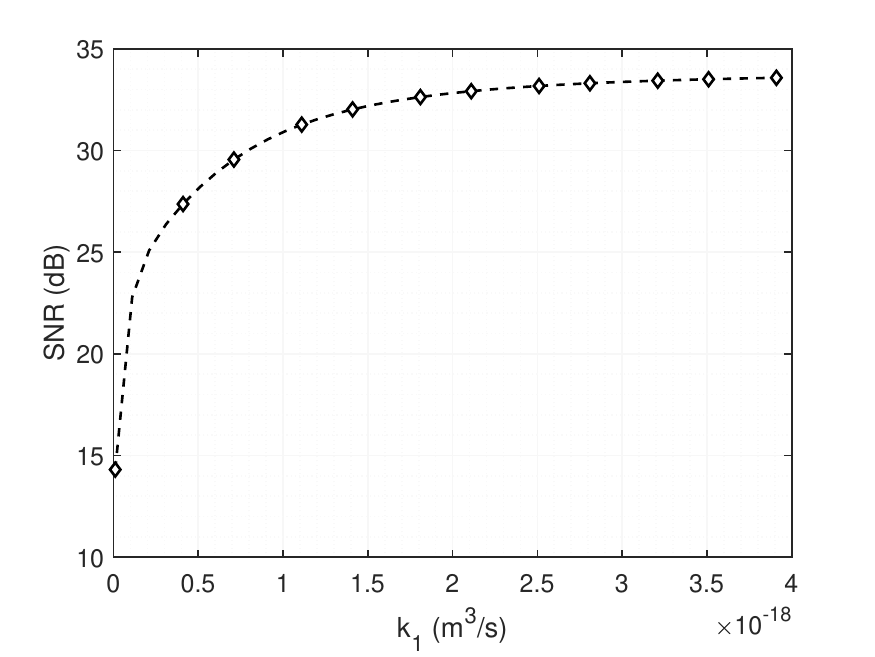}}\\[-1ex]
\subfloat[]{\includegraphics[width=0.24\textwidth]{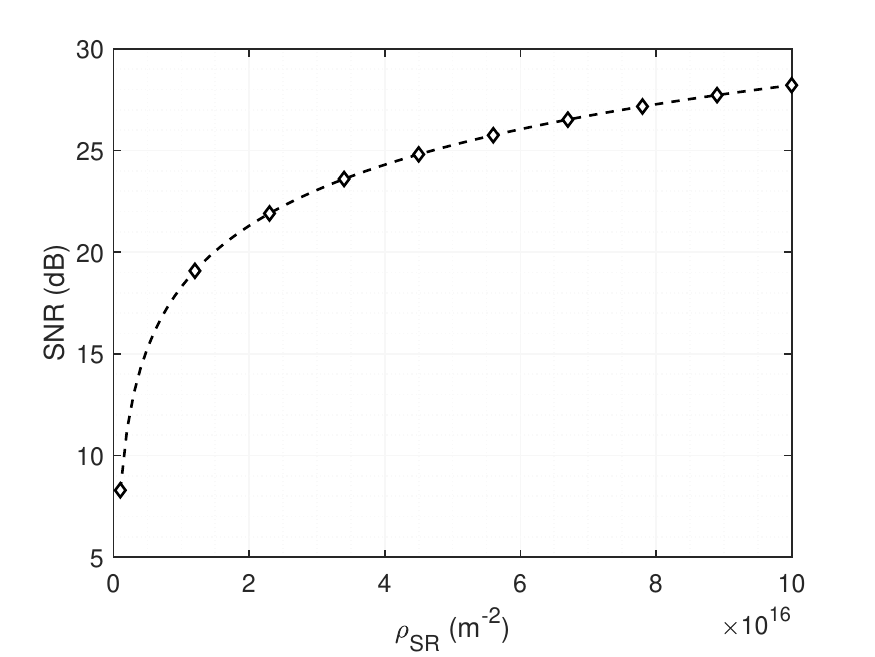}}
\hfil
\subfloat[]{\includegraphics[width=0.24\textwidth]{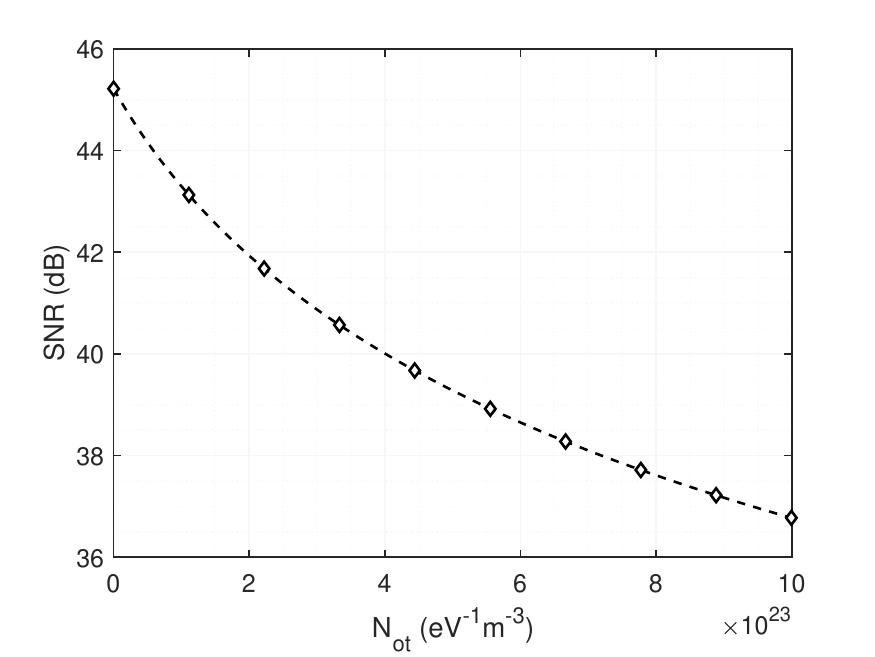}}
\hfil
\subfloat[]{\includegraphics[width=0.24\textwidth]{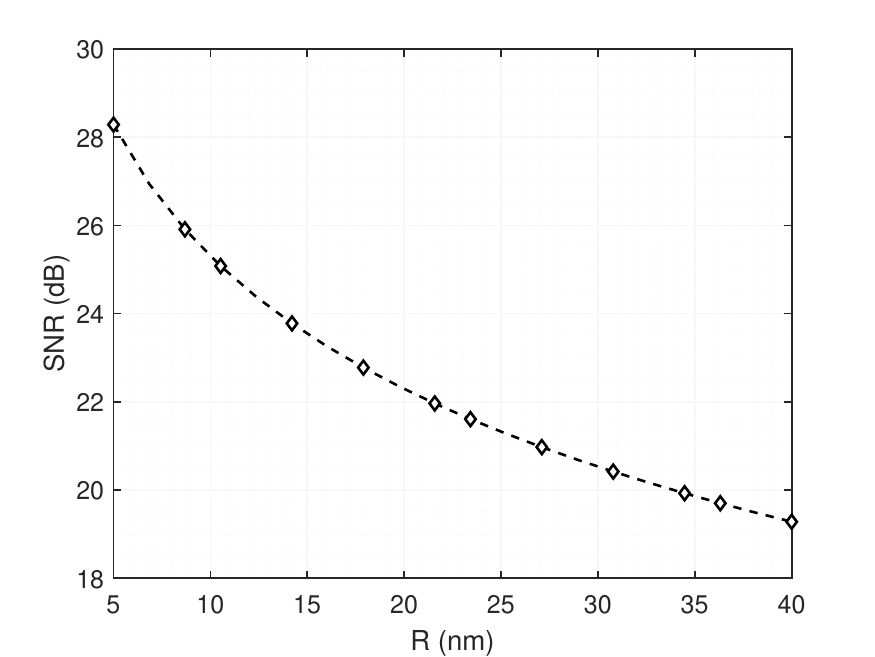}}
\caption{SNR as a function of (a) number of transmitted ligands $N_m$, (b) Tx-Rx distance $d$, (c) average flow velocity $u$, (d) intrinsic binding rate $k_1$, (e) surface receptor concentration $\rho_{SR}$, (f) oxide trap density $N_{ot}$, and (g) nanowire radius $R$.}
\label{fig9}
\end{figure*}

\subsection{Receiver Noise}

The contributions of binding and flicker noise to the overall noise spectrum are illustrated in Fig.~\ref{noise}. As discussed previously, thermomechanical noise is excluded since its influence on total noise power is negligible. Flicker noise dominates at very low ($f \ll 0.01~\text{Hz}$) and very high ($f \gg 100~\text{Hz}$) frequencies, whereas binding noise prevails in the intermediate frequency range. The results in Fig.~\ref{noise} are obtained using the default simulation parameters and remain consistent with our previous studies~\cite{aktas2021mechanical, aktas2022weight}, where binding noise becomes more pronounced at lower ligand concentrations due to the reduction in correlation time. Fig.~\ref{noise3d} illustrates the total output current noise PSD ($S_{I_m}$) of the cylindrical array-based Flexure-FET MC-Rx as a function of frequency $f$ and the number of transmitted ligands $N_m$.

\subsection{Signal-to-Noise Ratio}

The output SNR performance of the cylindrical array-based Flexure-FET MC-Rx is analyzed under various system, molecular, and receiver design parameters. The output SNR for symbol $m$ is expressed as \cite{kuscu2016modeling},

\begin{equation}\label{eq:19}
SNR_{out,m} = \frac{\mu_{I_m}^2}{\sigma_{I_m}^2},
\end{equation}
where $\mu_{I_m} = I_{DS1}/\mu_{S_m}$ represents the mean output current, and $\sigma_{I_m}^2$ is the current variance defined in (\ref{eq:15}) and (\ref{eq:var}).

As shown in Fig.~\ref{fig9}(a), the output SNR of the array-based cylindrical Flexure-FET MC-Rx increases with the number of transmitted ligands $N_m$ due to the stronger induced current at the receiver. However, the improvement saturates once the receptor sites become occupied, limiting additional molecular binding events~\cite{kuscu2016modeling}. Similarly, as illustrated in Fig.~\ref{fig9}(b), the SNR decreases with larger transmitter--receiver distance $d$ because of diffusion losses and channel attenuation. 

Figure~\ref{fig9}(c) illustrates the influence of the average flow velocity $u$ on the output SNR. Higher flow velocities accelerate molecular transport within the microchannel, resulting in more efficient ligand delivery to the receiver and reduced signal attenuation. Consequently, the SNR improves as $u$ increases, reflecting the enhanced mass transport efficiency under stronger flow conditions.

Receiver design parameters also influence the SNR characteristics. As shown in Fig.~\ref{fig9}(e), increasing the surface receptor concentration $\rho_{SR}$ enhances the probability of ligand binding, leading to a stronger mechanical response at the receiver. However, beyond a certain density, additional receptors provide diminishing returns, as the surface becomes saturated and steric hindrance limits further binding efficiency. In contrast, Fig.~\ref{fig9}(f) shows that increasing the oxide trap density $N_{ot}$ negatively affects SNR by amplifying low-frequency flicker noise (\ref{eq:16}). A higher $N_{ot}$ introduces additional fluctuations in the transducer’s mechanical–electrical coupling, which appear as increased noise in the output current.

As illustrated in Fig.~\ref{fig9}(g), increasing the nanowire radius $R$ results in a gradual decrease in SNR. Larger radii yield stiffer nanowires, which exhibit smaller mechanical deflections under molecular binding forces, thereby lowering the transduction efficiency. The selected range of $R$ values (5–40 nm) is consistent with reported silicon nanowire dimensions used in FET-based biosensors (10–100 nm)~\cite{tintelott2021process}. These observations are in agreement with the trends reported for the earlier planar Flexure-FET MC-Rx design~\cite{aktas2021mechanical}, indicating that the cylindrical array-based structure preserves the same SNR behavior under comparable system conditions.

\begin{figure*}[!t]
\centering
\subfloat[]{\includegraphics[width=0.48\columnwidth]{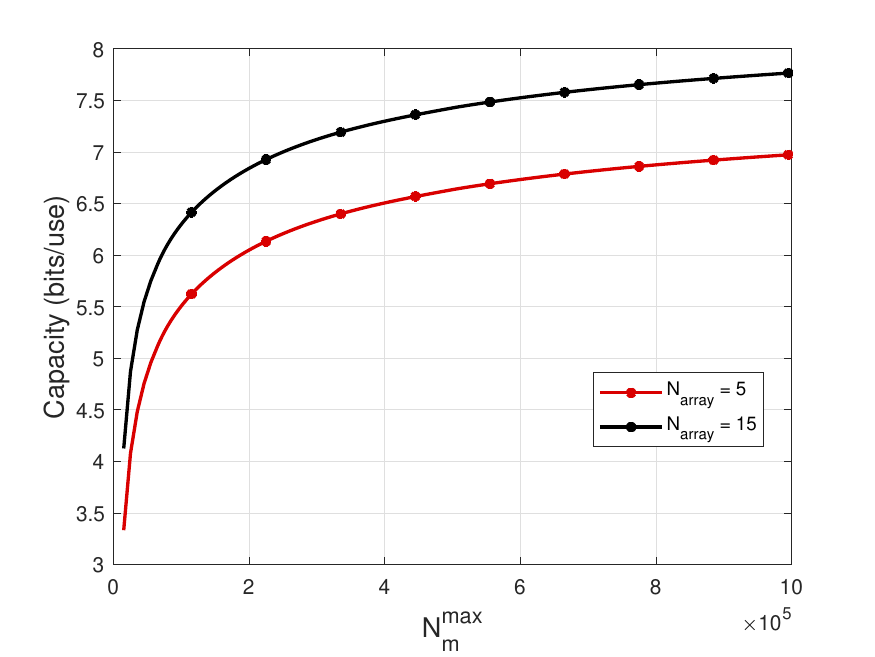}%
\label{fig_case_a}}
\hfil
\subfloat[]{\includegraphics[width=0.48\columnwidth]{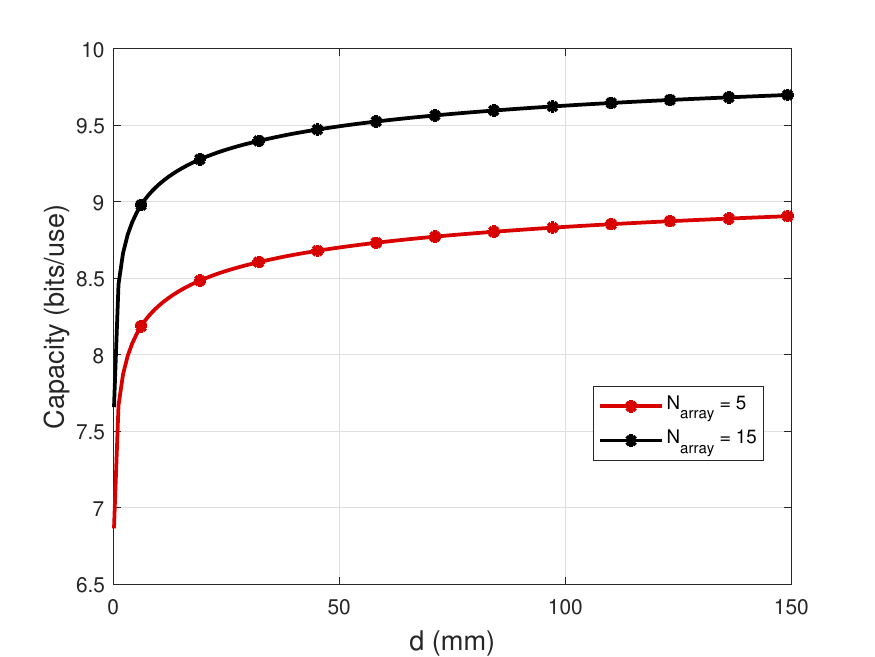}%
\label{fig_case_b}}
\hfil
\subfloat[]{\includegraphics[width=0.48\columnwidth]{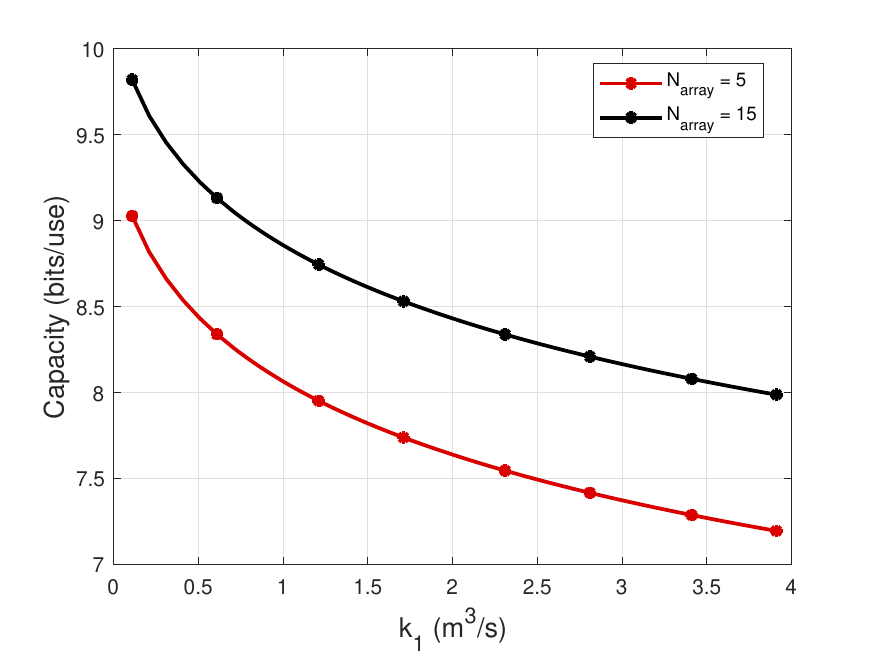}%
\label{fig_case_c}}
\hfil
\subfloat[]{\includegraphics[width=0.48\columnwidth]{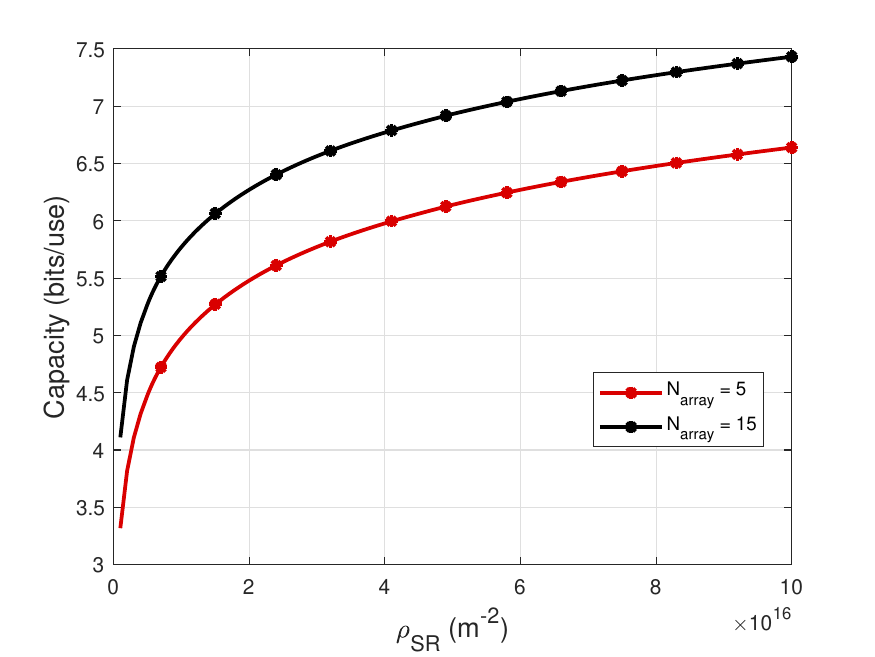}%
\label{fig_case_d}}
\caption{Channel capacity $C$ as a function of (a) maximum number of transmitted ligands $N_{tx}^{\max}$, (b) Tx-Rx distance $d$, (c) intrinsic binding rate $k_1$, and (d) surface receptor concentration $\rho_{SR}$.}

\label{fig10}
\end{figure*}

\subsection{Capacity}

The information transfer capability of the proposed MC-Rx is considered within the framework of a continuous-input, continuous-output memoryless channel. While the information is physically carried by a finite number of ligand molecules, the transmitted signal is treated as a continuous variable owing to its broad dynamic range \cite{kuscu2016capacity}.  
Accordingly, the closed-form expression for the channel capacity is adopted from the analytical derivation in \cite{kuscu2016capacity}, which establishes this relationship by maximizing the mutual information between the transmitted ligand signal and the receiver output under the assumption of Gaussian noise,

\begin{equation}
\begin{aligned}
C = & \frac{1}{2} \log_2\!\left(\frac{N_r}{2 \pi e}\right)
+ \log_2\!\Bigg[
\sin^{-1}\!\left(
L\,\frac{N_{tx}^{\max} - K_D / \beta_{ch}}
{N_{tx}^{\max} + K_D / \beta_{ch}}
\right)
\\
& \quad -
\sin^{-1}\!\left(
L\,\frac{N_{tx}^{\min} - K_D / \beta_{ch}}
{N_{tx}^{\min} + K_D / \beta_{ch}}
\right)
\Bigg],
\end{aligned}
\label{eq:capacity}
\end{equation}
where \(N_{tx}^{\min}\) and \(N_{tx}^{\max}\) denote the minimum and maximum number of transmitted ligands. 

The channel attenuation factor \(\beta_{ch}\) is defined in (\ref{eq:cons}).
The parameter \(L\) represents the effect of the receiver transduction characteristics and noise, defined as,
\begin{equation}
L =
\sqrt{
\frac{g_{\text{FET}}^2 \psi_L^2 N_r}
{4\sigma_F^2 + g_{\text{FET}}^2 \psi_L^2 N_r}
},
\end{equation}
where \(\sigma_F^2\) denotes the flicker noise variance, 
\begin{equation}\label{eq:16}
\sigma_F^2 = \int_{-\infty}^{\infty} S_{I_m^{F}}(f)\, df.
\end{equation}

Fig.~\ref{fig10}(a) shows the channel capacity as a function of the maximum number of transmitted ligands, \(N_m^{\max}\). The capacity increases with \(N_m^{\max}\) due to enhanced receptor occupancy and improved signal-to-noise ratio. The receiver with a larger nanowire array (\(N_{\text{array}} = 15\)) achieves higher capacity than the smaller one (\(N_{\text{array}} = 5\)) owing to its greater active surface area and stronger molecular transduction. Overall, increasing \(N_m^{\max}\) and array size improves information transfer until capacity saturation is reached.

In addition, Fig.~\ref{fig10}(b) shows the channel capacity versus transmitter–receiver distance \(d\). The capacity rises rapidly at short distances and then saturates as \(d\) increases. At small \(d\), the receiver experiences saturation due to excessive ligand concentration, reducing signal discrimination. As \(d\) grows, moderate concentrations improve sensitivity, but beyond a certain range, diffusion and advection losses limit capacity. The larger array maintains higher capacity owing to its greater molecular capture and transduction efficiency.

Furthermore, Fig.~\ref{fig10}(c) shows the channel capacity as a function of the ligand–receptor binding rate constant \(k_1\). The capacity decreases with increasing \(k_1\), since faster binding kinetics shorten the correlation time of receptor fluctuations, reducing the distinctiveness between molecular signal levels. At low \(k_1\), slower binding allows more stable receptor occupancy states, enhancing signal separability and thus capacity. The receiver with a larger nanowire array consistently achieves higher capacity due to its larger active area and improved transduction efficiency.

Finally, Fig.~\ref{fig10}(d) shows the channel capacity versus surface receptor concentration \(\rho_{SR}\). The capacity increases with \(\rho_{SR}\) due to improved ligand binding and higher SNR, but the growth saturates as receptor sites approach full occupancy. The larger array achieves higher capacity than the smaller one because of its greater sensing area and stronger transduction capability.

\section{Conclusion}

This paper presented a cylindrical nanowire array–based Flexure-FET MC-Rx as an extended and physically grounded architecture for MC systems. The proposed design enhances the Flexure-FET framework by incorporating a suspended nanowire array that enables distributed electromechanical coupling and increased design flexibility through multiple geometric parameters. An analytical end-to-end model was developed to characterize the electromechanical behavior, molecular interaction dynamics, noise characteristics, and information-theoretic performance of the receiver.

The results demonstrated how device-level factors, including nanowire geometry, receptor kinetics, and electromechanical response, and channel-related parameters, such as molecular transport and concentration, jointly determine system performance in terms of detection sensitivity, SNR, and channel capacity. Incorporating geometric degrees of freedom enhances the tunability of the receiver design, providing a versatile platform that can be adapted to advanced MC paradigms. In particular, the cylindrical array structure enables receptor-specific surface functionalization and spatial indexing, paving the way for mixture detection, multi-ligand sensing, and spatially modulated MC schemes.

Overall, the proposed architecture represents a significant step toward scalable and physically realizable MC receivers, bridging the gap between theoretical MC models and practical device implementations, and establishing a foundation for future OMC and IoNT applications.

\bibliographystyle{IEEEtran}
\bibliography{main}

\end{document}